\documentclass[letterpaper, 10 pt, journal]{IEEEtran} 

\IEEEoverridecommandlockouts                              
\usepackage{amsmath,mathrsfs}
\usepackage{amstext}
\usepackage{color,colortbl}
\usepackage[table]{xcolor}
\usepackage{amsfonts}
\usepackage{amssymb,amsbsy}
\usepackage{graphicx}
\usepackage{epstopdf}
\usepackage{url}
\usepackage[us]{datetime}
\usepackage{cite}
\usepackage[thmmarks, amsmath]{ntheorem}

\definecolor{Gray}{gray}{0.9}
\definecolor{LightCyan}{rgb}{0.88,1,1}
\definecolor{LightMagenta}{rgb}{1,0.88,1}
\definecolor{LightOrange}{rgb}{1,0.99,0.88}
\definecolor{LightGreen}{rgb}{0.9,1,0.8}
\definecolor{DarkOrange}{rgb}{0.98,0.91,0.71}
\definecolor{DarkGreen}{rgb}{0.67,0.88,0.69}

\allowdisplaybreaks[1]
\graphicspath{{../}, {../Bio/}, {../Fig/}, {../Fig02/}, {../Fig03/}, {../Fig04/}, {../Fig05/}, {../Fig06/}, {../Fig07/},
{../Fig08/}, {../Fig09/}, {../Fig10/}, {../Fig11/}, {../Fig12/}, {../Fig13/}}

\theorembodyfont{\normalfont}
\theoremseparator{:}
\newtheorem{claim}{Claim}
\newtheorem{remark}{Remark}
\newtheorem{proposition}{Proposition}
\theoremstyle{nonumberplain}
\theoremsymbol{$\square$}

\newcommand{\bfx}[0]{\mathbf{x}}
\newcommand{\bfy}[0]{\mathbf{y}}
\newcommand{\bfu}[0]{\mathbf{u}}
\newcommand{\bfA}[0]{\mathbf{A}}
\newcommand{\bfB}[0]{\mathbf{B}}
\newcommand{\bfI}[0]{\mathbf{I}}

\newcommand{\sat}[0]{\textrm{sat}}
\newcommand{\sign}[1]{\text{sign}\left(#1\right) }
\newcommand{\satfun}[1]{g\left(#1\right)}
\newcommand{\rmj}[0]{{\rm j}}
\newcommand{\diff}[0]{\textrm{d}}

\title{
A Nonlinear Car-following Controller Design Inspired By Human-driving Behaviors to Increase Comfort and Enhance Safety
}
\IEEEpubid{This paper is currently under peer review. This preprint is dedicated for REVIEW ONLY.}

\author{Wubing~B.~Qin
\thanks{Manuscript revised \currenttime, \today.%
}
\thanks{W.~B.~Qin is with the Department of Mechanical Engineering, University of Michigan, Ann Arbor, MI 48109, USA. (Email: wubing@umich.edu,).}%
}
\begin{document}
\maketitle
\begin{abstract}
This paper investigates the car-following problem and proposes a nonlinear controller that considers driving comfort, safety concerns, steady-state response and transient response. This controller is designed based on the demands of lower cost, faster response, increased comfort, enhanced safety and elevated extendability from the automotive industry.
Design insights and intuitions are provided in detail. Also, theoretical analysis are performed on plant stability, string stability and tracking performance of the closed-loop system. Conditions and guidelines are provided on the selection of control parameters. Comprehensive simulations are conducted to demonstrate the efficacy of the proposed controller in different driving scenarios.
\end{abstract}


\section{Introduction}

Recent decades have witnessed a growing interest in vehicle automation in academia and industry due to its potential in improving safety, mobility, fuel economy, and traffic throughput \cite{Marsden_2001, VanderWerfShladover02, Dav07, Askari_2016, Li_AAP_2017}. Vehicle automation involves implementation of many advanced driving features that can be categorized into longitudinal features, lateral features, side features and auxiliary features. These features have various levels of autonomy \cite{SAE_J3016_2016}, from advanced driver assistant systems (ADAS) to autonomous vehicles (AVs), and eventually to connected automated vehicles (CAVs).
Among the longitudinal features, longitudinal controller \cite{Huang_2012} for the car-following problem is the most important and has been studied extensively worldwide.

The early development on longitudinal controllers dates back to the 1960s \cite{LevineAthans66, Garrard_TR_1973, Chu_TS_1974, Chiu_JDSMC_1977, Olson_TVT_1979, Sklar_TVT_1979, Hauksdottir_TVT_1985}. Thereafter, different techniques are applied to the development of sensor-based control, often referred to as adaptive cruise control (ACC) \cite{Shladover_JDSMC_1978, IoaChi93, Rajamani_02, Winner_2012, Mattas_2021}.
From the beginning of this century, the booming wireless communication technology has fostered the development of communication-enhanced control techniques, which utilize vehicle-to-everything (V2X) communication to supplement the information that is not accessible to onboard sensors. These techniques include cooperative adaptive cruise control (CACC) \cite{Naus10, DesChai_ITS_2011, Milanes_Shladover_2014, Ploeg14, ShlNowLuFer15, Shengbo16} and connected cruise control (CCC) \cite{Ge_Orz_TRC_2014, Orosz16}, which have been demonstrated to perform well in experiments.

Concurrently, the automotive industry is dedicated to the practical deployment of these advanced driving features onto production vehicles. As a result, AVs are easily overloaded with tens of tasks, such as
ACC, lane centering (LC), forward collision avoidance (FCA), evasive steering assist (ESA), etc. These algorithms are implemented on middleware modules that bind the upper-level planning/decision-making modules with the lower-level actuator control modules. For cost reduction, middleware modules on production vehicles are usually affordable but less powerful micro-controllers.
Therefore, the automotive industry sets extremely high requirements on longitudinal controller for level-2-plus (L2+) vehicles.

The first requirement is that the controller must be computationally inexpensive and implementable on middleware modules.
Many existing longitudinal controllers require online optimization \cite{Akhegaonkar_2018, Chen_ITS_2019, Jia_IFAC_2018, Mamouei_TRC_2018, Plessen_TCST_2018} such as model predictive control (MPC), dynamic programming, etc. To make this type of controllers fit middleware platforms in real-time implementations, it is a common practice to shorten prediction horizon, downgrade numerical precision, and linearize models. The resulting performance degradation hinders the deployment of online optimization technique on production vehicles.

The second requirement is that the longitudinal controller must take driving comfort and safety into consideration. On the one hand, in a dynamic traffic environment, the controller must ensure natural behaviors (i.e., reasonable acceleration and jerk) of the ego vehicle when scenario changes due to cut-in, cut-out, merging, splitting, etc. On the other hand, the ego vehicle must be able to avoid imminent collisions. Although optimization-based control algorithms can include these aspects as constraints, solving constrained optimal control problem is beyond the capability of middleware modules. Whereas, it is rather difficult to consider driving comfort in other control techniques, such as LQR, gain scheduling, etc.

\IEEEpubidadjcol

The third requirement is that the controller must be maintainable and extendable. Due to the fact that most longitudinal controllers in literature cannot meet the aforementioned requirements, currently the automotive industry utilizes another control technique on level-1 (L1) or level-2 (L2) vehicles, that is, lookup tables (LUTs). However, its major issue in maintainability and extendability hinders its deployment onto L2+ vehicles because of the following reasons. To characterize scenario-dependent actions, the state space is partitioned into small regions to form a multi-dimensional LUT, which depends on range, range rate, desired time headway, ego vehicle speed, etc. This easily leads to explosion in the number of tuning parameters. For example, to obtain a coarse LUT, thousands of parameters need to be determined in field experiments.
Due to this cumbersome tuning process, these LUTs are then fixed as the baseline design that cannot be changed thereafter. The common solution to resolve issues identified afterwards is to overlay patches, which makes the algorithm unmaintainable. Consequently, LUT-based algorithms are not extendable for increased level of automation.

In this paper, we investigate the car-following problem and attempt to design a longitudinal controller to meet the requirements of the automotive industry for L2+ vehicles. Due to the computational limitations, we reinvestigate the widely-used linear controller \cite{CYLiang_VSD_1999, Santhanakrishnan_ITS_2003, Guvenc_CSM_2006, Cook_TAC_2007, Bu_2012, diBernardo15, Lee_TRC_2017, GCDC_Halmstad_2018}, and reaffirmed its effectiveness and stability performance around the uniform flow equilibrium. However, the lack of consideration on transient response generates unnatural driving behaviors and deteriorates driving comfort.
Therefore, we propose a nonlinear controller by extending this linear controller in the following way: 1) we ensure the topological equivalence between the linear controller and the proposed nonlinear controller around the uniform flow equilibrium; 2) we improve driving comfort in transient phase by applying closed-form nonlinear functions to mimic human-driving behaviors; 3) we add a feedforward term to avoid collisions in safety-concerned scenarios. The proposed controller is computationally cheap with increased comfort, enhanced safety, and elevated extendibility.

The major contributions are as follows. Firstly, transient response is included and can be guaranteed in the design, while the existing controllers typically leave the task of ensuring transient response to planning algorithms. Thus, the proposed controller has much similar behaviors to human drivers while approaching the desired uniform flow equilibrium when the initial speed error or range error is large. Secondly, it has enhanced safety in two aspects. On the one hand, in safety-concerned scenarios imminent collisions will be avoided by the feedforward term in the design.
On the other hand, due to the deliberate design of transient response, the proposed controller will not overreact in case of temporary failures/malfunctions. For example, there is a common problem that perception algorithm may detect non-existing ``ghost" vehicles, or misclassify far-away stationary objects (trees, traffic signs, etc.) as stationary vehicles. Although these false detections only appear for few seconds, existing controllers typically generate harsh brakes because of large speed difference with respect to the ``ghost" vehicle regardless of range. These unexpected harsh brakes deteriorate driving comfort and pose a severe threat to safety for following vehicles. In contrast, when a ``ghost" vehicle appears far away, the proposed controller may start with coasting and then re-accelerate upon its disappearance. The guarantee on transient response ensures that the actions are subtle throughout this process without noticeable behaviors. Another major contribution is that the parametrization of driving comfort and physical limits makes the proposed controller extendible to more scenarios, and integrable with upper-level algorithms for L2+ vehicles.

This paper is organized in the following way. In Section~\ref{sec:ctrl}, we start with the preliminaries on the car-following problem, control design objectives and dynamic models. Then we propose a nonlinear controller and explain it in detail. In Section~\ref{sec:analysis}, we derive closed-loop nonlinear dynamics, and analyze its stability and tracking performance. The conditions on plant stability and string stability, and guidelines on tracking performance are given on the selection of control gains. In Section~\ref{sec:sim_res}, we conduct simulations to validate the proposed controller in multiple typical driving scenarios and compare the results against those using a widely-used controller. Finally, conclusions are drawn, and future research directions are pointed out in Section~\ref{sec:conclusion}.

\section{Controller Design\label{sec:ctrl}}

In this section we start with the problem statement and present the nonlinear controller in Section \ref{sec:ctrl_design}. Then the design details on the nonlinear feedback law and collision-free feedforward law are provided in Section~\ref{sec:fdbck_ctrl} and \ref{sec:collision_avoid}, respectively.

\subsection{Problem Statement and Nonlinear Controller Design \label{sec:ctrl_design}}
\begin{figure}
  \centering
  \includegraphics[scale=0.6]{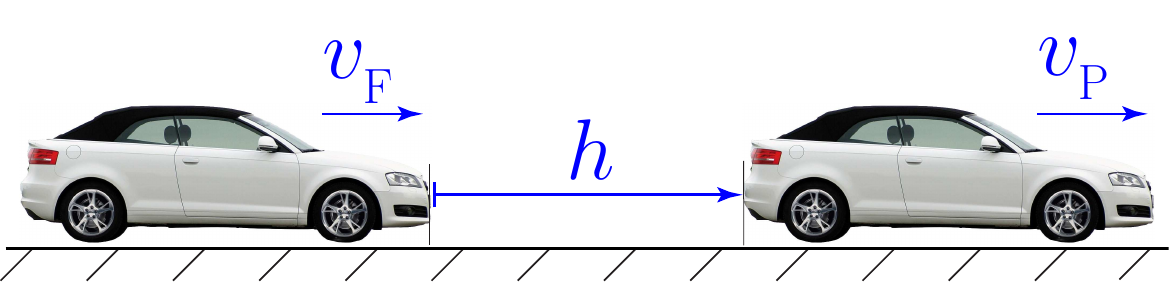}\\
  \caption{Car-following scenario.}\label{fig:car_follow}
\end{figure}

In this paper, we consider the predecessor-follower pair shown in Fig.~\ref{fig:car_follow}, and start with the simplest model, that is,
\begin{equation}\label{eqn:toy_model}
  \begin{split}
    \dot{h} & = v_{\rm P} -v_{\rm F}\,,\\
    \dot{v}_{\rm F} & = a_{\rm des}\,,
  \end{split}
\end{equation}
where $v_{\rm P}$, $v_{\rm F}$ and $h$ represent the predecessor speed, the follower speed and the inter-vehicle distance, respectively. Also, $a_{\rm des}$ is the desired acceleration given by the car-following controller. This model implicitly indicates that the follower acceleration
is assumed to be capable of tracking the desired acceleration perfectly, which is non-realistic in practice. However, for the controller we propose in this paper, it provides insights and facilitates understanding. For more realistic scenarios, we refer readers to Section~\ref{sec:extension} on how to extend the proposed controller to models with higher fidelity.

We assume that the follower can utilize either onboard sensors or V2V communication to obtain the predecessor speed $v_{\rm P}$, follower speed $v_{\rm F}$ and inter-vehicle distance $h$.
The objective is to design a controller for the follower in this car-following scenario that can generate the desired acceleration $a_{\rm des}$ based on the accessible information and other customizable user preferences.
This controller must meet the following objectives:
\begin{enumerate}
  \item The \emph{uniform flow equilibrium} is stabilizable at steady state, that is, the follower eventually equates its speed with the predecessor while maintaining the desired distance given by the so-called \emph{range policy}.
  \item When the initial state is far away from the uniform flow equilibrium, the follower must respond reasonably in the transient phase while approaching the predecessor and settling down at the uniform flow equilibrium.
  \item In case of emergency that is within the follower's physical capability, the follower is able to avoid collision.
\end{enumerate}
Regarding the range policy, we use the constant time-headway policy in this paper, i.e.,
\begin{align}\label{eqn:hdes}
  h_{\rm des} &= h_{0} + v_{\rm P}\, t_{\rm h} \,,
\end{align}
where $h_{0}$ is the standstill distance, and $t_{\rm h}$ is the desired time-headway. We remark that this desired distance is based on the predecessor speed instead of the follower speed that is widely used in literature, and the reason will be explained in Section~\ref{sec:sim_res}.

\begin{figure}
  \centering
  \includegraphics[scale=1]{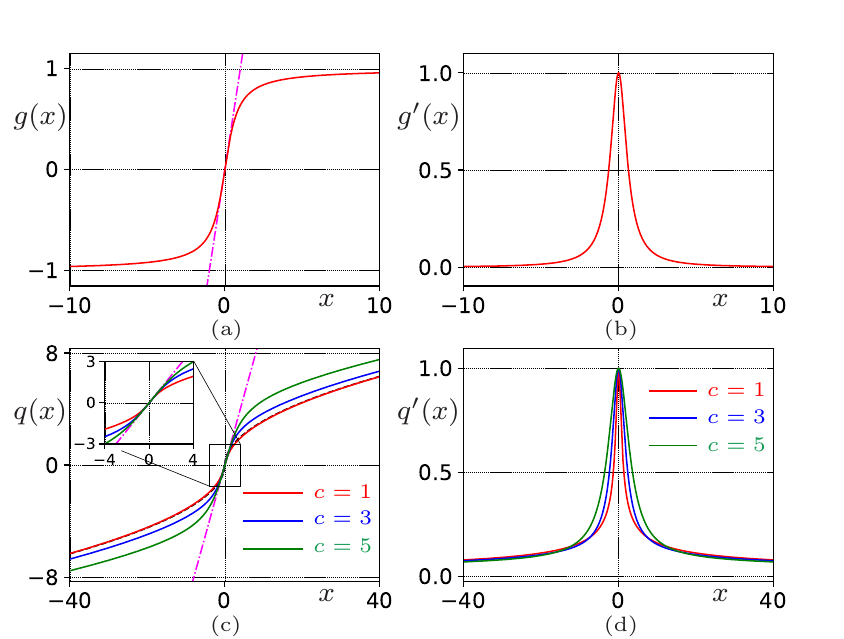}\\
  \caption{(a, b) Function $g(x)$ and its derivative $g'(x)$. (c, d) Function $q(x)$ and its derivative $q'(x)$ for different $c$ value when $b=0.5$.}\label{fig:wrappers}
\end{figure}

We propose the following controller
\begin{align}\label{eqn:ades_nonlin}
  a_{\rm des} & = a_{\rm cf} + a_{\rm fb}\,,
\end{align}
that consists of a feedback term $a_{\rm fb}$ to guarantee reasonable steady state response (objective 1) and transient response (objective 2), and a feedforward term $a_{\rm cf}$ to avoid collision (objective 3).
By defining the errors from the desired state as
\begin{align}\label{eqn:delta_v_h}
  \hat{v} &=v_{\rm P}-v_{\rm F}\,, &
  \hat{h} & = h - h_{\rm des}\,,
\end{align}
we design the collision-free feedforward law
\begin{align}\label{eqn:ff_cf_term}
  a_{\rm cf} &=  \max\left\{-\dfrac{\hat{v}^{2}\cdot H(-\hat{v})}{2\,\max\{h-h_{\min},\, \varepsilon\} }\,,\; a_{\min} \right\}\,,
\end{align}
where $H(x)$ is the heaviside step function, $h_{\min}$ is the minimum allowed inter-vehicle distance, $\varepsilon>0$ is used to avoid singularity, and $a_{\min}<0$ is the physical minimum acceleration.
The feedback term
\begin{align}\label{eqn:fb_all}
  a_{\rm fb} &= \bar{a}_{\rm fb}+a_{\sat}\cdot\satfun{\tfrac{k_{1} S}{a_{\sat}}}\,,
\end{align}
consists of a feedback law that ensures the follower state evolves along a desired surface $S=0$ explained below in the transient phase, and the corresponding acceleration $\bar{a}_{\rm fb}$.
Here $k_{1}$ is a control gain, $a_{\sat}>0$ is the maximum allowed acceleration, and $g(x)$ denotes a wrapper function of $x$ satisfying the following properties:
\begin{enumerate}
  \item It is continuously differentiable and strictly increasing over $\mathbb{R}$.
  \item It is an odd function, i.e., ${g(x)=-g(-x)}$ for ${x\in \mathbb{R}}$.
  \item It is bounded in $[-1, \, 1]$, i.e., $g:\mathbb{R}\to [-1, \, 1]$.
  \item Its derivative strictly decreases for ${x\in \mathbb{R}_{\geq0}}$ such that ${g'(0)=1}$ and ${\lim\limits_{x \to +\infty}  g'(x) = 0}$.
\end{enumerate}
In this paper we use the smooth wrapper function
\begin{equation}
    g(x) =\tfrac{2}{\pi}\arctan \Big(\tfrac{\pi}{2} x\Big)\,, \label{eqn:satfunction}
\end{equation}
and Fig.~\ref{fig:wrappers}(a, b) plots this function and its derivative.

Inspired by human-driving behaviors, we design the surface
\begin{align}
  \widehat{S} & =\hat{v} + q\left(k_{2}\, \hat{h};\, \frac{a_{\rm com}}{k_{2}}\right)\,,\label{eqn:surf_S0}\\
  S &= \max\{\min\{\widehat{S},\, v_{\max} -v_{\rm F}\}, \,-v_{\rm F}\}\,,\label{eqn:surf_S}
\end{align}
to ensure the transient response and steady-state response, where $k_{2}$ is another control gain, $a_{\rm com}$ is the comfortable acceleration in the transient phase, $v_{\rm max}$ is the preset maximum speed, and $q(x; \, b)$ denotes a shaping function with a parameter $b>0$ satisfying the following properties:
\begin{enumerate}
  \item It is continuously differentiable and strictly increasing over $\mathbb{R}$.
  \item It is odd in $x$, i.e., ${q(x)=-q(-x)}$ for ${x\in \mathbb{R}}$.
  \item It has a curvilinear asymptote $y=\sqrt{2\,b\, x}$ as $x\to +\infty$, i.e., $ \lim\limits_{x \to +\infty}  \left\{q(x)-\sqrt{2\,b\, x}\right\} = 0$.
  \item Its derivative is continuous and strictly decreasing over ${\mathbb{R}_{\geq0}}$ and $q'(0)=1$.
\end{enumerate}
Note that here we use the shorthand notation $q(x)$ to represent $q(x;\,b)$ when highlighting parameters is not necessary. This notation will be maintained for other functions as well throughout the paper.
Henceforth, the underlying acceleration indicated by the surface \eqref{eqn:surf_S0} is
\begin{align}\label{eqn:fb_ades_S}
  \bar{a}_{\rm fb} & = q'\left(k_{2}\, \hat{h};\tfrac{a_{\rm com}}{k_{2}}\right)k_{2}\hat{v}\,,
\end{align}
In this paper, we use the smooth shaping function
\begin{align}\label{eqn:wrapper_q}
  q(x; b) &= g(\tfrac{x}{c})\sqrt{2\,b\, x\, g(\tfrac{x}{c})+c^{2}}\,,
\end{align}
where $c>0$ is a slackness parameter and $g(x)$ is the wrapper function \eqref{eqn:satfunction}. Fig.~\ref{fig:wrappers}(c, d) shows the function and its derivative.

\subsection{Feedback Control\label{sec:fdbck_ctrl}}
To gain more insights, we start with the simple scenario where the predecessor speed is constant ($\dot{v}_{\rm P}\equiv 0$), but the initial errors ($\hat{v}$ and $\hat{h}$) can be either large or small. One may refer to Section~\ref{sec:analysis} for more details on the dynamic tracking performance when the predecessor speed varies ($\dot{v}_{\rm P}\neq 0$). Also, we put aside the collision-free feedforward term in this part to facilitate understanding.


The nonlinear feedback control law \eqref{eqn:fb_all} aims to ensure that the follower can adjust speed reasonably based on the errors $\hat{v}$ and $\hat{h}$ regardless of their magnitude. For now we neglect the feedforward term \eqref{eqn:ff_cf_term}, and assume that the saturation limits in \eqref{eqn:surf_S} are not reached. When the errors $\hat{v}$ and $\hat{h}$ are small, one may obtain the linearized controller
\begin{align}\label{eqn:ades_lin}
  a_{\rm des}& = a_{\rm fb} = \bar{a}_{\rm fb}+k_{1}S \,,
\end{align}
where
\begin{align}\label{eqn:lin_ctrl_fb_terms}
  \bar{a}_{\rm fb} & = k_{2}\hat{v}\,, &
  S  & =\hat{v} +q^{0}(\hat{h})\,,
\end{align}
and
\begin{align}\label{eqn:lin_ctrl_q0}
  q^{0}(x) & = k_{2}x\,.
\end{align}
This is equivalent to the widely-used linear feedback controller
\begin{align}\label{eqn:ctrl_lin_0}
  a_{\rm fb} & = \bar{k}_{1}\hat{v} +\bar{k}_{2}\hat{h}\,,
\end{align}
in the literature \cite{CYLiang_VSD_1999, VanderWerfShladover02, Santhanakrishnan_ITS_2003, Guvenc_CSM_2006, Cook_TAC_2007, Bu_2012, diBernardo15, Lee_TRC_2017, GCDC_Halmstad_2018}, where
\begin{align}
  \bar{k}_{1} & = (k_{1} +k_{2})\,, & \bar{k}_{2} &=k_{1}k_{2}\,.
\end{align}
Many other existing controllers also apply a similar technique that utilizes a linear combination on multiple error terms. This technique has been proven to be effective when initial errors are small. However, in the following we show that linear combination of error terms produces unexpected behaviors in the car-following problem when initial errors are large.

Utilizing $S $ and $\hat{v}$ as the states, we can rewrite the closed-loop system (\ref{eqn:toy_model}, \ref{eqn:ctrl_lin_0}) into
\begin{align}
  \dot{S} & = -k_{1}S \,,\label{eqn:lin_s_dv_model_1}\\
  \dot{\hat{v}} &= -k_{1}S -k_{2}\hat{v}\,.\label{eqn:lin_s_dv_model_2}
\end{align}
by using the assumption that $\dot{v}_{\rm P}\equiv 0$. It is easy to verify that if $k_{1}>0$ and $k_{2}>0$,
\begin{align}\label{eqn:lin_s_dv_equil}
  S^{\ast}&=0\,, &
  \hat{v}^{\ast} &=0\,,
\end{align}
is a stable equilibrium of the closed-loop system (\ref{eqn:lin_s_dv_model_1}, \ref{eqn:lin_s_dv_model_2}), which corresponds to the desired uniform flow equilibrium. On the one hand, \eqref{eqn:lin_s_dv_model_1} governs the evolution of transient response while approaching the steady-state equilibrium \eqref{eqn:lin_s_dv_equil}. On the other hand, $S $ converges to zero faster than $\hat{v}$ since $S$ acts as an excitation input to \eqref{eqn:lin_s_dv_model_2}. Therefore, the closed-loop system evolves along the surface $S \approx 0$ in the transient phase, and gradually settles down to the equilibrium \eqref{eqn:lin_s_dv_equil}. This observation will be verified by simulations in Section~\ref{sec:sim_res}.
By differentiating $S = 0$, one can obtain the underlying acceleration $\bar{a}_{\rm fb}$ (cf.~\eqref{eqn:lin_ctrl_fb_terms}) in order to evolve along this surface.

Utilizing \eqref{eqn:delta_v_h}, we can rewrite the second term in the controller \eqref{eqn:ades_lin} into
\begin{align}\label{eqn:a_fb_0_track_S0}
  \hat{a}_{\rm fb} &:=k_{1}S  = k_{1}(v_{\rm des}-v_{\rm F})\,,
\end{align}
where
\begin{align}\label{eqn:vdes_0_lin}
  v_{\rm des} &= v_{\rm P}+k_{2}\hat{h}\,.
\end{align}
This implies that the linearized controller attempts to make the vehicle speed $v_{\rm F}$ follow the desired speed $v_{\rm des}$, which is based on the predecessor speed and range difference against desired value.
We will show this interpretation in Section~\ref{sec:sim_res} with more details.

Indeed, this linearized controller \eqref{eqn:ctrl_lin_0} can achieve reasonable performance when initial errors $\hat{v}$ and $\hat{h}$ are small enough. However, investigating the underlying acceleration $\bar{a}_{\rm fb}$ (cf.~\eqref{eqn:lin_ctrl_fb_terms}) to evolve along the surface $S =0$ when initial errors are large, we observe the following unnatural behaviors:
\begin{itemize}
  \item When the predecessor is slower ($\hat{v}<0$), the follower will always decelerate despite the large inter-vehicle distance ($\hat{h}>0$), since $\bar{a}_{\rm fb}$ only depends on speed difference. Thus, a harsh brake may be generated unexpectedly in the case of large speed difference ($\hat{v}\ll 0$) even when the predecessor is far away ($\hat{h}\gg 0$).
  \item Similarly, when the predecessor is faster ($\hat{v}>0$), the follower will always accelerate despite the collision-imminent inter-vehicle distance ($\hat{h}\ll 0$).
\end{itemize}

To resolve these issues, we investigate the natural responses of human drivers in the following scenarios:
\begin{itemize}
  \item When the follower approaches a slow-moving predecessor far away ($\hat{v}<0$ and $\hat{h} > 0$), human drivers tend to decelerate at a near-constant comfortable acceleration $-a_{\rm com}<0$ to minimize jerk and increase comfort. Ideally, the follower speed equals the predecessor speed when reaching the desired range, i.e., $\hat{v}=0$ and $\hat{h} = 0$. By applying principles of kinematics on the relative motion of the predecessor and follower, one can approximately characterize this tendency with
        \begin{align}
          \hat{v}^{2} = 2\,a_{\rm com}\hat{h} \quad \Longrightarrow \quad
          \hat{v}+\sqrt{2\,a_{\rm com}\hat{h}}=0\,.
        \end{align}

  \item Similarly, when the follower closely follows a fast-moving predecessor ($\hat{v}>0$ and $\hat{h} < 0$), human drivers tend to accelerate at a constant comfortable acceleration $a_{\rm com}$ such that its speed equals the predecessor speed when reaching the desired range, i.e., $\hat{v}=0$ and $\hat{h} = 0$. This tendency can be approximately characterized by
        \begin{align}
          \hat{v}^{2} = -2a_{\rm com}\hat{h} \quad \Longrightarrow \quad
          \hat{v}-\sqrt{-2\,a_{\rm com}\hat{h}}=0\,.
        \end{align}
\end{itemize}
In summary, when initial errors are large, the surface that human drivers tend to follow is approximately
\begin{align}\label{eqn:surf_human}
  \hat{v}+\hat{q}(\hat{h})=0\,,
\end{align}
where
\begin{align}\label{eqn:q0_human}
  \hat{q}(x)&=\sign{x}\sqrt{2\,a_{\rm com}x\,\sign{x}}\,.
\end{align}
This tendency of constant acceleration approach is only valid for large initial errors. The approximation error will produce undesired behaviors when errors become small. This is because the desired acceleration given by \eqref{eqn:q0_human} changes abruptly from $a_{\rm com}$ to $0$ upon reaching the desired equilibrium, leading to unexpected jerky behaviors.

Therefore, we combine the applicable scenarios of linear control and human-driver-inspired control into the surface \eqref{eqn:surf_S0}, and requires that $q(x)$ is approximately equal to $q^{0}(x)$ when $|x|$ is small (property 4), but $\hat{q}(x)$ when $|x|$ is large (property 3). Indeed, one can use any $q(x)$ satisfying the aforementioned properties to obtain reasonable performance. In this paper, we choose \eqref{eqn:wrapper_q} by replacing non-continuous sign function with a smooth wrapper function and making relevant modifications to satisfy these properties. Fig.~\ref{fig:wrappers}(c, d) shows this function $q(x)$ and its corresponding derivative $q'(x)$ when the slackness parameter equals $1, \, 3,\, 5$, respectively. As indicated, larger $c$ value implies a moderate decrease of acceleration from $b$ to $0$ that resolves the issue of jerky behaviors. However, larger $c$ value also results in a wider effective range of linear strategy that may deteriorate performance. We observe that larger $c$ values satisfying $c<2b$ achieve good performance.

To evolve along the desired surface while approaching the predecessor from large initial errors, the underlying acceleration $\bar{a}_{\rm fb}$ can be obtained by differentiating $\widehat{S}=0$; cf.~\eqref{eqn:fb_ades_S}.
Similar to \eqref{eqn:a_fb_0_track_S0}, we can rewrite $\widehat{S}$ in \eqref{eqn:surf_S0} into
\begin{align}\label{eqn:surf_S0_interp}
  \widehat{S} & = \hat{v}_{\rm des}-v_{\rm F}
\end{align}
where the desired speed is
\begin{align}
  \hat{v}_{\rm des} &= v_{\rm P}+q(k_{2} \hat{h})\,.
\end{align}
To resolve the issue that this desired speed might be outside the follower speed range, we update it to
\begin{align}\label{eqn:vdes_nonlin}
 v_{\rm des} = \max\{\min\{\hat{v}_{\rm des}, \,v_{\max}\}, \,0\}\,
\end{align}
which is equivalent to \eqref{eqn:surf_S}.

In the linear controller \eqref{eqn:a_fb_0_track_S0}, the same control gain $k_{1}$ is applied for both small and large errors of $S $ in order to track the desired speed \eqref{eqn:vdes_0_lin}. However, practically we prefer larger gains for small errors to achieve better tracking performance, but smaller gains for large errors to avoid ``overreaction" and potential oscillations. To resolve this conflict, gain scheduling are typically utilized such that different gains can be applied for errors in different ranges. However, this technique requires strenuous tuning and stitching to ensure satisfactory performance. Alternatively, we utilize a new method that can effectively decrease the gain when necessary. In particular, the wrapper function $g(x)$ is designed for this purpose. As indicated in Fig.~\ref{fig:wrappers}(a, b), the derivative of the wrapper function $g(x)$ monotonically decreases with respect to $|x|$. As a result, the nonlinear controller \eqref{eqn:fb_all} utilizing this wrapper function will decrease the gain accordingly based on the error magnitude.

One may notice that we have only investigated two typical scenarios ($\hat{v}\cdot\hat{h} <0$) when designing the controller. Indeed, it can be verified that if $k_{1}>0$, the proposed controller performs well in other scenarios ($\hat{v}\cdot\hat{h} >0$) as well:
i) when the follower follows a close-and-slow-moving predecessor ($\hat{v}<0$ and $\hat{h} < 0$), (\ref{eqn:surf_S0}, \ref{eqn:surf_S}) yields that $S<0$ and the controller \eqref{eqn:fb_all} decelerates the follower.
ii) when the follower approaches a far-and-fast-moving predecessor ($\hat{v}>0$ and $\hat{h} > 0$), (\ref{eqn:surf_S0}, \ref{eqn:surf_S}) yields that $S>0$ and the controller \eqref{eqn:fb_all} accelerates the follower.

\begin{remark}\label{remark:ctrl_extension}
  This nonlinear controller is extended from the widely-used controller \eqref{eqn:ctrl_lin_0} by resolving the issue of unnatural behaviors in the transient phase when the follower states are far away from the uniform flow equilibrium. However, when the follower states are close enough to the uniform flow equilibrium, the controller \eqref{eqn:ctrl_lin_0} linearly approximates the proposed nonlinear controller.
\end{remark}

\subsection{Collision Avoidance\label{sec:collision_avoid}}
The feedback law ensures that the follower approaches the predecessor along the desired surface in the transient phase and settles down to the desired equilibrium at steady state. However, it cannot guarantee that the deceleration is large enough to avoid collision. In practice, a collision is only possible when preceding vehicle is slower, i.e., $\hat{v} <0$. To avoid collision, the minimum collision-free deceleration $-\hat{a}_{\rm cf}<0$ can be applied such that the follower speed equals the predecessor speed upon reaching the minimum inter-vehicle distance $h_{\min}$, yielding
\begin{align}
  \hat{v}^{2} = -2\,\hat{a}_{\rm cf} (h-h_{\min}) \enskip \Longrightarrow \enskip
  \hat{a}_{\rm cf} =-\dfrac{\hat{v}^{2}}{2\, (h-h_{\min})}\,,
\end{align}
by applying principles of kinematics on the relative motion.
Then we update $\hat{a}_{\rm cf}$ into  $a_{\rm cf}$ given in \eqref{eqn:ff_cf_term} by taking the following aspects into consideration: (1) This term is only needed when collision is possible, i.e., $\hat{v}<0$; (2) The actual distance might temporarily be slightly shorter than $h_{\min}$ due to response delay; (3) The resulting deceleration might exceed vehicular physical limit $a_{\min}$ and potentially cause permanent damage to the braking system.

\section{Analysis\label{sec:analysis}}
There are three key aspects in car-following design that need to be investigated. The first one is whether the closed-loop system possesses a \emph{uniform flow equilibrium} representing the follower's capability of traveling at the same speed as the predecessor while maintaining desired distance. The second one relates to stabilizability of this equilibrium, often referred to as \emph{plant stability} or \emph{internal stability}. It reflects the follower's ability to reach the desired equilibrium when the predecessor speed is constant. The third aspect is the so-called \emph{string stability} \cite{Peppard_TAC_1974, SwaHed96, Cook_SCL_2005, Studli_SM17, Besselink_2017}, which represents the follower's capability to attenuate fluctuations imposed on the predecessor speed. Research [\citen{Orosz_Wilson_2009}, \citen{OroWilSte10}] shows that many phantom traffic jams are typically caused by string unstable vehicles.

In this section, we will investigate these aspects of the closed-loop system (\ref{eqn:toy_model}) with controller (\ref{eqn:ades_nonlin}-\ref{eqn:wrapper_q}). Also, we will investigate the tracking performance along the specifically designed surface $S$. For simplicity, stability and tracking performance are analyzed based on the corresponding linearized model, because it is topologically equivalent to the nonlinear system in the neighborhood of the equilibrium. We derive conditions on plant stability and string stability, and provide guidelines on how to achieve best tracking.

To study tracking performance in the transient phase, the nonlinear transformation (\ref{eqn:delta_v_h}, \ref{eqn:surf_S0}, \ref{eqn:surf_S}) is applied to the closed-loop system (\ref{eqn:toy_model}, \ref{eqn:ades_nonlin}-\ref{eqn:wrapper_q}) to transform states from $(h, \, v_{\rm F})$ to $(\hat{S},\, \hat{v})$, whose inverse transformation is
\begin{equation}\label{eqn:non_transf_inv}
 \begin{split}
  h&=h_{\rm des}+q^{-1}(S-\hat{v})\,, \\
  v_{\rm F} &= v_{\rm P}-\hat{v}\,,
 \end{split}
\end{equation}
where $q^{-1}(x)$ represents the inverse function of $q(x)$.
Assuming that the saturation limits in \eqref{eqn:surf_S} are not reached, one can obtain the closed-loop system
\begin{equation}
\begin{split}\label{eqn:nonlin_s_dv_model}
  \dot{S} & = -a_{\rm cf}-g(k_{1}S)+\bigr(1-k_{2}t_{\rm h}Q(S-\hat{v})\bigr) \dot{v}_{\rm P}\,,\\
  \dot{\hat{v}} &= -a_{\rm cf} -g(k_{1}S)-Q(S-\hat{v})k_{2}\hat{v}+\dot{v}_{\rm P}\,,
\end{split}
\end{equation}
where
\begin{align}
  Q(x) &=q'\left(q^{-1}(x)\right)\,.
\end{align}

It can be easily verified that when the predecessor speed is constant, i.e., $v_{\rm P}\equiv v_{0}$, the closed-loop system (\ref{eqn:nonlin_s_dv_model}) possesses the uniform flow equilibrium
\begin{align}\label{eqn:equil_vL_const}
  S^{\ast} &=0\,,&
  \hat{v}^{\ast} &=0\,,
\end{align}
where the follower travels at the same speed as the predecessor and maintains desired distance given by its range policy.


When the predecessor speed varies around a constant nominal speed $v_{0}$, we can define the input perturbation as
\begin{align}
  \tilde{v}_{\rm P}&= v_{\rm P}-v_{0}\,,&
\end{align}
and the resulting perturbations on states as
\begin{align}
  \tilde{S} &= S-S^{\ast}\,,&
  \tilde{v} &= \hat{v}-\hat{v}^{\ast}\,.
\end{align}
By defining the state, the input and the output as
\begin{align}
  \bfx &=
  \begin{bmatrix}
    \tilde{S} \\ \tilde{v}
  \end{bmatrix}\,,&
  \bfu &= \dot{\tilde{v}}_{\rm P}\,,&
  \bfy & =
  \begin{bmatrix}
    \tilde{S} \\ \tilde{v}
  \end{bmatrix}\,,
\end{align}
and utilizing the inverse function theorem, one can linearize (\ref{eqn:nonlin_s_dv_model}) about the equilibrium \eqref{eqn:equil_vL_const} and obtain
\begin{equation}\label{eqn:lin_ss_model}
  \begin{split}
      \dot{\bfx} &=\bfA \bfx+\bfB\bfu\,,\\
      \bfy &=\bfx\,,
  \end{split}
\end{equation}
where
\begin{align}
  \bfA &=
  \begin{bmatrix}
    -k_{1} & 0\\
    -k_{1} & -k_{2}
  \end{bmatrix},\,&
  \bfB &=
  \begin{bmatrix}
    1-k_{2}t_{\rm h}\\
    1
  \end{bmatrix}.
\end{align}
The characteristic equation is
\begin{align}\label{eqn:char_eqn}
  \det (s\, \bfI-\bfA)&= s^{2}+(k_{1}+k_{2})s+k_{1}k_{2}=0\,,
\end{align}
where $s\in \mathbb{C}$ is the eigenvalue of the system. To ensure stability, all eigenvalues must lie in the left half of the complex plane.
\begin{proposition}\label{prop:plant_stable_condition}
  The closed-loop system is plant stable if 
  \begin{align}\label{eqn:plant_stab_cond}
      k_{1}&>0\,,&
      k_{2} &>0\,.
    \end{align}
\end{proposition}
\begin{IEEEproof}
    Applying Routh-Hurwitz criterion to \eqref{eqn:char_eqn}, one can obtain the necessary and sufficient condition of the linearized system \eqref{eqn:lin_ss_model}, that is,
        \begin{align}
      k_{1}+k_{2} &>0\,,&
      k_{1}k_{2} &>0\,,
    \end{align}
    which is equivalent to \eqref{eqn:plant_stab_cond}.
    Moreover, the stability of the nonlinear system \eqref{eqn:nonlin_s_dv_model} is topologically equivalent to its linearized model \eqref{eqn:lin_ss_model} around the equilibrium.
\end{IEEEproof}

To study tracking performance and string stability, we define the transfer functions
\begin{align}
  \widetilde{G}_{1}(s) &=\dfrac{\widetilde{\mathcal{S}}(s)}{\widetilde{\mathcal{V}}_{\rm P}(s)}\,,&
  \widetilde{G}_{2}(s) &=\dfrac{\widetilde{\mathcal{V}}(s)}{\widetilde{\mathcal{V}}_{\rm P}(s)}\,,
\end{align}
where $\widetilde{\mathcal{V}}_{\rm P}(s)$, $\widetilde{\mathcal{V}}(s)$ and $\widetilde{\mathcal{S}}(s)$ are the Laplace transform of $\tilde{v}_{\rm P}(t)$, $\tilde{v}(t)$ and $\tilde{S}(t)$, respectively.
One can obtain these transfer functions as
\begin{align}
  \widetilde{G}_{1}(s) &=\dfrac{(1-k_{2}t_{\rm h})s(s+k_{2})}{s^{2}+(k_{1}+k_{2})s+k_{1}k_{2}}\,,\\
  \widetilde{G}_{2}(s) &=\dfrac{s(s+k_{1}k_{2}t_{\rm h})}{s^{2}+(k_{1}+k_{2})s+k_{1}k_{2}}\,,
\end{align}
through
\begin{align}
  \begin{bmatrix}
    \widetilde{G}_{1}(S) \\ \widetilde{G}_{2}(s)
  \end{bmatrix}&= s(s\,\bfI-\bfA)^{-1}\bfB\,.
\end{align}
Notice that
\begin{align}
  \widetilde{G}_{2}(s)&=\dfrac{\widetilde{\mathcal{V}}(s)}{\widetilde{\mathcal{V}}_{\rm P}(s)}= \dfrac{\widetilde{\mathcal{V}}_{\rm P}(s)-\widetilde{\mathcal{V}}_{\rm F}(s)}{\widetilde{\mathcal{V}}_{\rm P}(s)}=1-\dfrac{\widetilde{\mathcal{V}}_{\rm F}(s)}{\widetilde{\mathcal{V}}_{\rm P}(s)}\,,
\end{align}
where $\widetilde{\mathcal{V}}_{\rm F}(s)$ is the Laplace transform of $\tilde{v}_{\rm F}=v_{\rm F}-v_{0}$.
Thus, the transfer function $G(s)$ from the perturbations on predecessor speed to those on follower speed can be obtained, that is,
\begin{equation}
\begin{split}
  G(s)&:=\dfrac{\widetilde{\mathcal{V}}_{\rm F}(s)}{\widetilde{\mathcal{V}}_{\rm P}(s)}=1-\widetilde{G}_{2}(s)\\
  &=\dfrac{(k_{1}+k_{2}-k_{1}k_{2}t_{\rm h})s+k_{1}k_{2}}{s^{2}+(k_{1}+k_{2})s+k_{1}k_{2}}\,.
\end{split}
\end{equation}
By defining the corresponding magnitude as
\begin{align}
  \widetilde{M}_{1}(\omega)&=|\widetilde{G}_{1}(\rmj\, \omega)|\,,&
  M(\omega)&=|G(\rmj\, \omega)|\,,
\end{align}
we obtain
\begin{align}
\widetilde{M}_{1}(\omega)&=\sqrt{\dfrac{(1-k_{2}t_{\rm h})^{2}\omega^{2}(k_{2}^{2}+\omega^{2})}{(k_{1}k_{2}-\omega^{2})^{2}+(k_{1}+k_{2})^{2}\omega^{2}}}\,,\\
M(\omega)&=\sqrt{\dfrac{(k_{1}+k_{2}-k_{1}k_{2}t_{\rm h})^{2}\omega^{2}+k_{1}^{2}k_{2}^{2}}{(k_{1}k_{2}-\omega^{2})^{2}+(k_{1}+k_{2})^{2}\omega^{2}}}\,.
\end{align}
Based on Fourier theory, a periodic signal can be represented as a countable sum of sines and cosines, which can also be extended to absolutely integrable non-periodic signals using Fourier transform. With the assumption that the perturbations on the predecessor speed are absolutely integrable and have finite energy, string stability can be characterized by attenuation of sinusoidal perturbations at all frequencies
according to the superposition principle. Thus, string stability condition requires that the amplification
ratio is always less than 1 at all excitation frequencies, that is,
\begin{align}
  \forall{\omega>0}:\  M(\omega)<1\,.\label{eqn:string_stab_cond_gen}
\end{align}
We remark that $M(0)=1$ always holds because the follower is able to follow the constant predecessor speed, which can be viewed as zero frequency excitations.

\begin{proposition}\label{prop:string_stable_condition}
  The closed-loop system is string stable if condition \eqref{eqn:plant_stab_cond} and the condition
  \begin{align}\label{eqn:string_stab_cond}
  k_{1}t_{\rm h}(k_{2}-2)\le 2(k_{2}t_{\rm h}-1)\,
\end{align}
hold.
\end{proposition}
\begin{IEEEproof}
Condition \eqref{eqn:string_stab_cond_gen} is equivalent to
\begin{align}
  \forall \omega>0: \ P(\omega)&<0\,,
\end{align}
where $P(\omega)$ is the difference between the numerator and denominator of $M^{2}(\omega)$. That is,
\begin{equation}
  \begin{split}
    P(\omega)&=-\omega^{4}+k_{1}k_{2}\big(k_{1}k_{2}t_{\rm h}-2(k_{1}+k_{2})t_{\rm h}+2\big)\omega^{2}\,,
  \end{split}
\end{equation}
yielding the condition
\begin{align}
 k_{1}k_{2}\big(k_{1}k_{2}t_{\rm h}-2(k_{1}+k_{2})t_{\rm h}+2\big)\le 0\,,
\end{align}
which is equivalent to \eqref{eqn:string_stab_cond} by considering the prerequisite condition \eqref{eqn:plant_stab_cond} on plant stability.
\end{IEEEproof}

\begin{figure}
  \centering
  \includegraphics[scale=1]{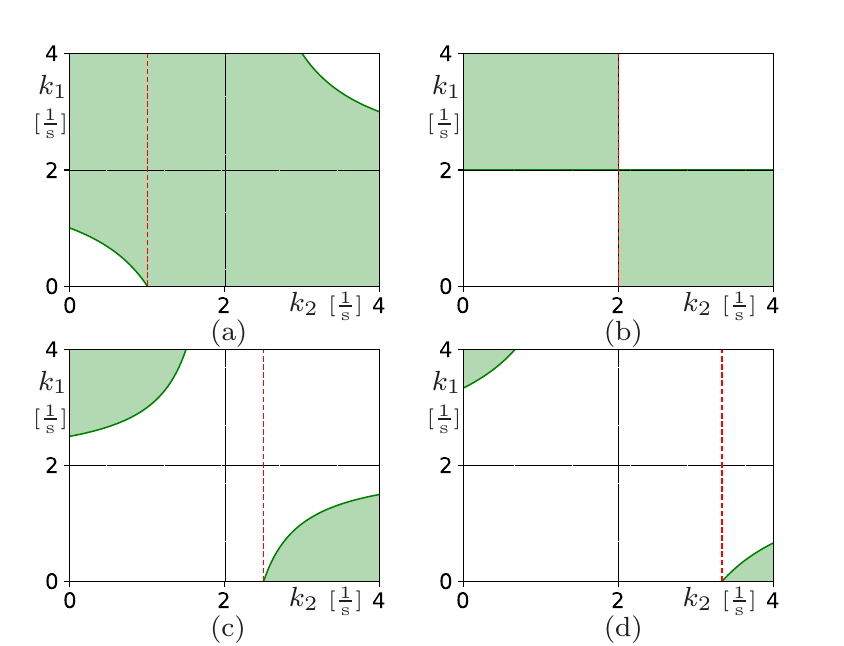}\\
  \caption{Stability diagrams. (a) $t_{\rm h}=1$ [s]. (b) $t_{\rm h}=0.5$ [s]. (c) $t_{\rm h}=0.4$ [s]. (d) $t_{\rm h}=0.2$ [s].\label{fig:stab_diag}}
\end{figure}
Fig.~\ref{fig:stab_diag} shows the string stability diagram in $(k_{2},\, k_{1})$-plane for different $t_{\rm h}$. The shaded regions represent string stable regions satisfying conditions (\ref{eqn:plant_stab_cond}, \ref{eqn:string_stab_cond}). Observe that the string stable regions shrink as the desired time headway $t_{\rm h}$ decreases.

Plant stability have ensured that when the predecessor speed is constant, the follower state will evolve along the desired surface $S=0$ to approach the uniform flow equilibrium \eqref{eqn:equil_vL_const}. When the predecessor speed varies, we also prefer good tracking performance about this surface to avoid unexpected behaviors caused by large tracking errors.
\begin{claim}\label{claim:tracking}
  To achieve good tracking along the designed surface $S=0$ while approaching equilibrium \eqref{eqn:equil_vL_const}, $k_{2}$ may be chosen in the neighborhood of $k_{2}^{\ast}=\frac{1}{t_{\rm h}}\,$.
\end{claim}
\begin{IEEEproof}
  One can calculate
  \begin{align}
    \dfrac{\diff}{\diff\omega}\widetilde{M}_{1}^{2}(\omega)&=\dfrac{2\, k_{1}^{2}(1-k_{2}t_{\rm h})^{2}\omega(k_{2}^{2}+\omega^{2})^{2}}{\bigr((k_{1}k_{2}-\omega^{2})^{2}+(k_{1}+k_{2})^{2}\omega^{2}\bigr)^{2}}\ge 0
  \end{align}
  for all $\omega\ge 0$, implying that $\widetilde{M}_{1}^{2}(\omega)$ increases monotonically for $\omega\in \mathbb{R}_{\ge 0 }$. Notice that $\widetilde{M}_{1}(\omega)$ is an even function, and
  \begin{align}
    \lim_{\omega\to +\infty}\widetilde{M}_{1}(\omega)& =|1-k_{2}t_{\rm h}|\,,
  \end{align}
  yielding that
  \begin{align}
    \forall \omega \in \mathbb{R}:\ \widetilde{M}_{1}(\omega) \le |1-k_{2}t_{\rm h}|\,.
  \end{align}
  Thus, choosing $k_{2}$ in the neighborhood of $k_{2}^{\ast}$ can minimize the upper bound of $\widetilde{M}_{1}(\omega)$, leading to the improvement in the tracking performance.
\end{IEEEproof}

In Fig.~\ref{fig:stab_diag}, the red dashed lines represent the preferred choices of $k_{2}=k_{2}^{\ast}$. One can observe that when the desired time headway $t_{\rm h}\ge 0.5$ [s], it is possible that the proposed controller can guarantee plant stability, string stability and zero tracking errors concurrently. However, when the desired time headway is too small, the condition on tracking performance needs to yield to the conditions on plant stability and string stability.

\begin{remark}
We derive the conditions in Proposition \ref{prop:plant_stable_condition} and \ref{prop:string_stable_condition} to demonstrate that in the parameter space there exists plant stable and string stable regions for the closed-loop system. However, these conditions are pretty general since we did not consider the effects of disturbances, time delays, actuator dynamics, digital control, optimality, etc. As mentioned in Remark \ref{remark:ctrl_extension}, the widely-used controller \eqref{eqn:ctrl_lin_0} linearly approximates the proposed nonlinear controller when the follower states are close enough to the uniform flow equilibrium. Also, performance degradation from these effects has been studied extensively for the controller \eqref{eqn:ctrl_lin_0}. Thus, previous results on the linear controller \eqref{eqn:ctrl_lin_0} are still applicable to the proposed nonlinear controller in the neighborhood of the equilibrium. We expect that the plant stable and string stable regions in the parameter space will shrink in general as these effects become more and more significant. In the worst scenario, these regions will disappear, implying that there are no parameters that can guarantee plant stability and string stability.
\end{remark}

\section{Simulation Results \label{sec:sim_res}}

In this section, we conduct numerical simulations to demonstrate the effectiveness of the proposed nonlinear car-following controller and discuss its extensions. In Section~\ref{sec:sim_const_speed_lead} we compare the performance of the proposed nonlinear controller against the widely-used linear controller when the follower approaches a constant-speed predecessor with large initial errors from the desired uniform flow equilibrium. In Section~\ref{sec:sim_vary_speed_lead} we demonstrate the performance of the proposed controller when the predecessor speed varies. This includes validation on full stops and string stability. In Section~\ref{sec:extension}, we show how the proposed controller can be extended to vehicle models with higher fidelity and how to handle uncertainties. The parameters used in the simulations are provided in Table.~\ref{tab:params}.

\begin{table}[!t]
\begin{center}
\renewcommand{\arraystretch}{1.3}
\rowcolors{1}{LightCyan}{LightMagenta}
\begin{tabular}{l|c|l}
\hline\hline
 \rowcolor{Gray}  Parameter & Value & Description\\
 \hline 
 $h_{0}$ [m]& $5$ & standstill distance\\
 $t_{\rm h}$ [s]& $1$ & desired time headway\\
 $h_{\rm min}$ [m]& $5$ & minimum allowed distance\\
 $\varepsilon$ [m]& $0.5$ & small value to avoid singularity\\
 $v_{\max}$ [m/s]& $35$ & driver preset maximum speed\\
 $c$ [m/s]& $1$ & slackness parameter\\
 $a_{\sat}$ [m/s$^{2}$] & $4$ & maximum allowed acceleration\\
 $a_{\min}$ [m/s$^{2}$] & $-10$ & physical minimum acceleration\\
 $a_{\rm com}$ [m/s$^{2}$] & $0.5$ & user-specific comfortable acceleration\\
 $k_{1}$ [s$^{-1}$]& $1.5$ & control gain\\
 $k_{2}$ [s$^{-1}$]& $1$ & control gain\\
 $k_{\rm I}$ [s$^{-1}$]& $0.1$ & control gain\\
 $\tau$ [s] & 0.8 & time constant of actuator dynamics\\
 $\Delta$ [m/s$^{2}$]& $0.5$ & disturbance of model \eqref{eqn:toy_model_disturb}\\
 $\tilde{\Delta}$ [m/s$^{2}$]& $0.5$ & disturbance of model \eqref{eqn:PT_model_disturb}\\
\hline\hline
\end{tabular}
\end{center}
\caption{Parameters used in the simulation. \label{tab:params}}
\end{table}

\subsection{Approaching Constant-speed Predecessor\label{sec:sim_const_speed_lead}}
\begin{figure}[!t]
  \centering
  \includegraphics[scale=1]{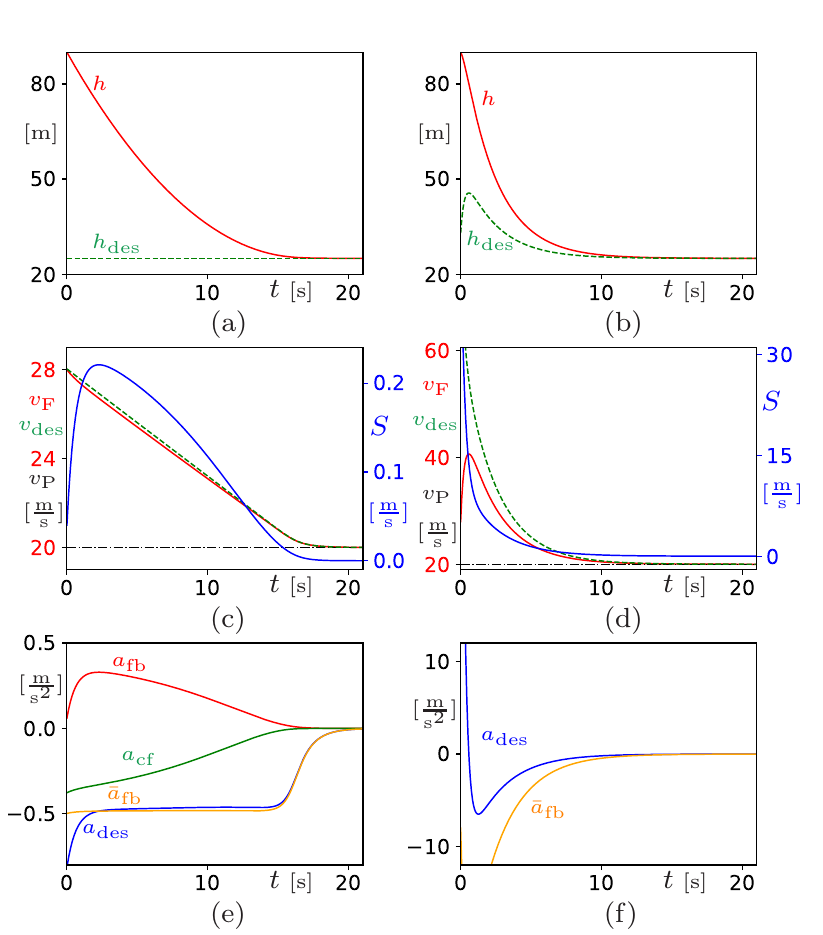}\\
  \caption{\label{fig:sim_lead_slow_far}The follower approaches a far-but-slow-moving predecessor. Left column: with the nonlinear controller (\ref{eqn:ades_nonlin}-\ref{eqn:wrapper_q}). Right column: with the linear controller \eqref{eqn:ctrl_lin_0}.}
\end{figure}

\begin{figure}[!t]
  \centering
  \includegraphics[scale=1]{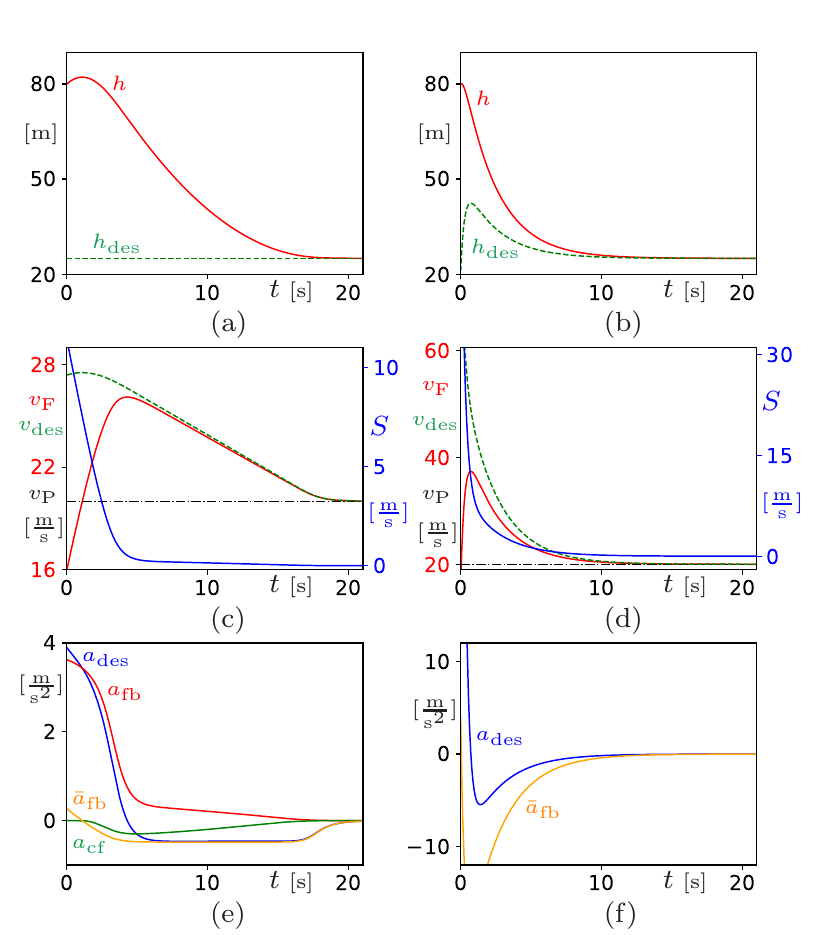}\\
  \caption{\label{fig:sim_lead_fast_far}The follower approaches a far-and-fast-moving predecessor. Left column: with the nonlinear controller (\ref{eqn:ades_nonlin}-\ref{eqn:wrapper_q}). Right column: with the linear controller \eqref{eqn:ctrl_lin_0}.}
\end{figure}

\begin{figure}[!t]
  \centering
  \includegraphics[scale=1]{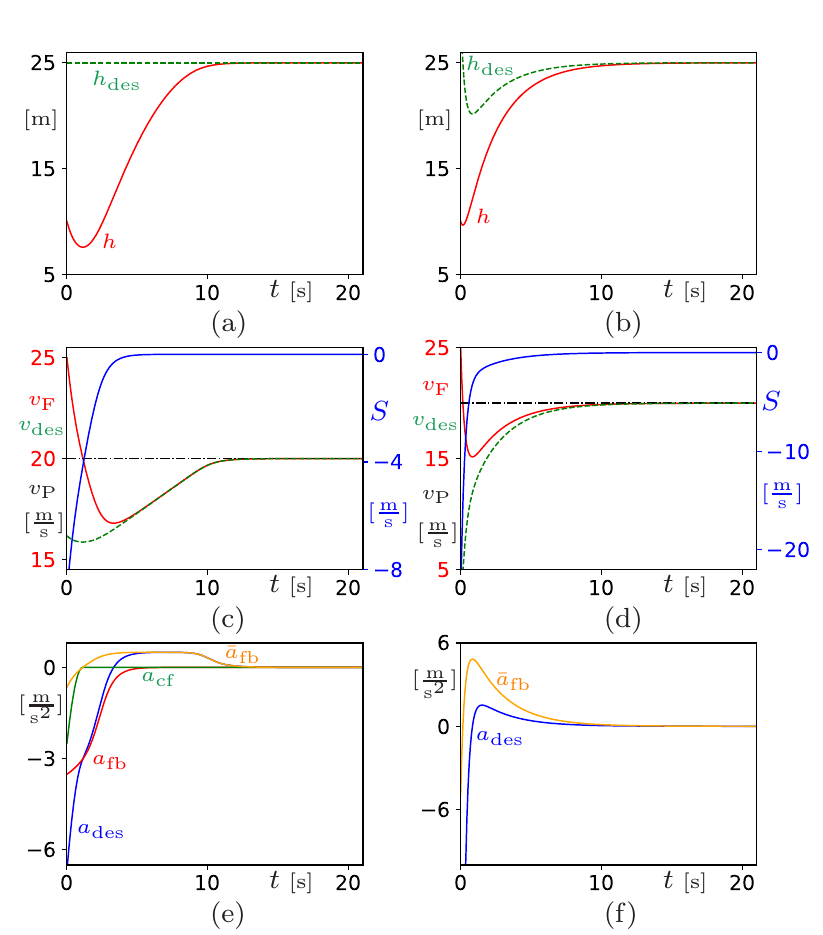}\\
  \caption{\label{fig:sim_lead_slow_close}The follower approaches a close-and-slow-moving predecessor. Left column: with the nonlinear controller (\ref{eqn:ades_nonlin}-\ref{eqn:wrapper_q}). Right column: with the linear controller \eqref{eqn:ctrl_lin_0}.}
\end{figure}

\begin{figure}[!t]
  \centering
  \includegraphics[scale=1]{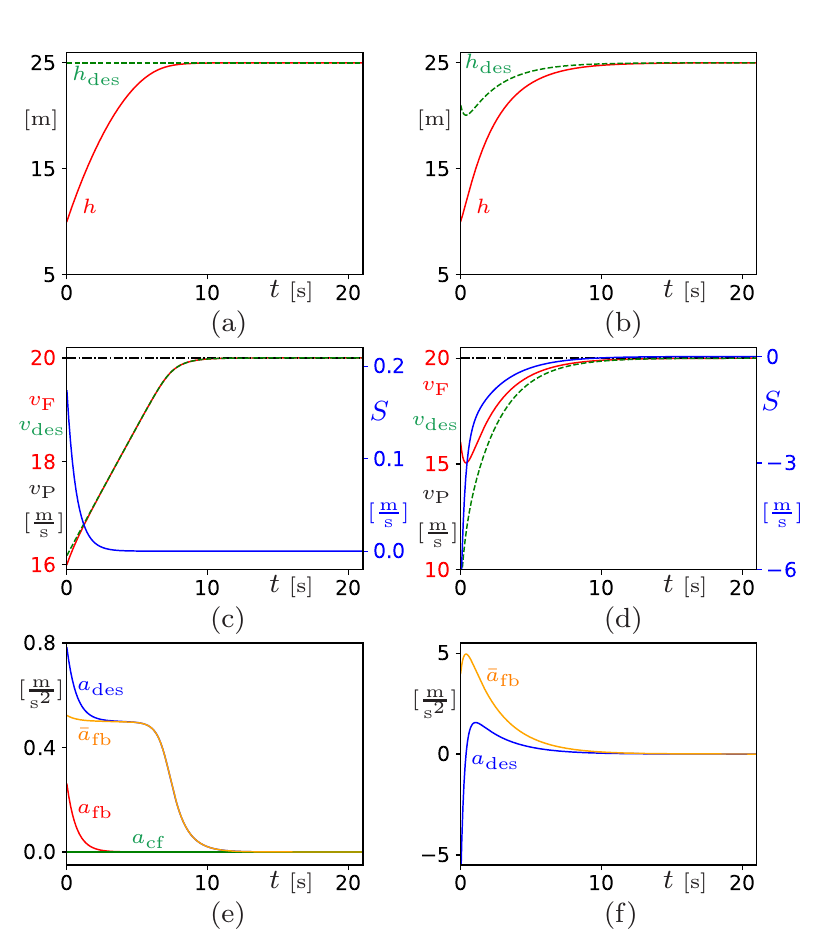}\\
  \caption{\label{fig:sim_lead_fast_close}The follower approaches a close-but-fast-moving predecessor. Left column: with the nonlinear controller (\ref{eqn:ades_nonlin}-\ref{eqn:wrapper_q}). Right column: with the linear controller \eqref{eqn:ctrl_lin_0}.}
\end{figure}

Figs.~\ref{fig:sim_lead_slow_far}-\ref{fig:sim_lead_fast_close} compare the performance of the proposed controller (\ref{eqn:ades_nonlin}-\ref{eqn:wrapper_q}) against the linear controller \eqref{eqn:ctrl_lin_0} in the scenarios where the follower approaches a far-but-slow-moving predecessor, far-and-fast-moving predecessor, close-and-slow-moving predecessor and close-but-fast-moving predecessor at constant speed, respectively. In particular, for the widely-used linear controller \eqref{eqn:ctrl_lin_0}, we change the range policy \eqref{eqn:hdes} to the more commonly-used one, that is,
\begin{align}\label{eqn:hdes_F}
  h_{\rm des} &= h_{0} + v_{\rm F}\, t_{\rm h} \,,
\end{align}
which is calculated based on the follower speed $v_{\rm F}$ instead of the predecessor speed $v_{\rm P}$. We remark that simulation results look similar for this linear controller with both range policies (\ref{eqn:hdes}, \ref{eqn:hdes_F}). When initial conditions are far from the uniform flow equilibrium, they always generate unexpected behaviors exhibited in these figures. Regarding the proposed nonlinear controller, the performance with range policy \eqref{eqn:hdes} is much better than that with range policy \eqref{eqn:hdes_F}. This is because the desired speed $v_{\rm des}$ in \eqref{eqn:vdes_nonlin} depends on the distance error $\hat{h}$ between the current value $h$ and the desired value $h_{\rm des}$. This desired distance $h_{\rm des}$ acts as ultimate desired value when the predecessor speed is not changing, because the ultimate desired speed of the follower is $v_{\rm P}$. Changing range policy \eqref{eqn:hdes} to \eqref{eqn:hdes_F} is equivalent to changing the ultimate desired distance to current desired distance that relies on current speed $v_{\rm F}$. Variations on current speed $v_{\rm F}$ in closed-loop control will in turn lead to variations in desired distance that deteriorate performance. Thus, in the following the proposed nonlinear controller are simulated with range policy \eqref{eqn:hdes}.

In these figures, the following layout and color scheme are maintained.
The left columns correspond to simulation results of the proposed nonlinear controller (\ref{eqn:ades_nonlin}-\ref{eqn:wrapper_q}), while the right columns correspond to the results of the linear controller \eqref{eqn:ctrl_lin_0}.
Also, the top, the middle and the bottom panels show the time profiles of the terms related to distance, speed and acceleration, respectively. Specifically, in panels (a, b), the red solid curves represent the inter-vehicle distance $h$, while the green dashed curves indicate the desired value $h_{\rm des}$ given by the range policy \eqref{eqn:hdes} or \eqref{eqn:hdes_F}. In panels (c, d), the black dot-dashed curves, the red solid curves and the green dashed curves represent the predecessor speed $v_{\rm P}$, the follower speed $v_{\rm F}$, and the desired speed $v_{\rm des}$ in \eqref{eqn:vdes_nonlin} or \eqref{eqn:vdes_0_lin}, respectively, while the blue solid curves using the vertical axis marked on the right side indicate the tracking error $S$ in \eqref{eqn:surf_S} or \eqref{eqn:lin_ctrl_fb_terms}. In panels (e, f), the blue, the red, the orange and the green solid curves represent the desired acceleration $a_{\rm des}$, the feedback term $a_{\rm fb}$, the underlying acceleration $\bar{a}_{\rm fb}$ in \eqref{eqn:fb_ades_S} or \eqref{eqn:lin_ctrl_fb_terms}, and the collision-free feedforward term $a_{\rm cf}$ in \eqref{eqn:ff_cf_term}, respectively. Note that different scales might be used to highlight the difference between the left columns and right columns.

Fig.~\ref{fig:sim_lead_slow_far} shows a scenario where the follower approaches a far-but-slow-moving predecessor with the initial conditions $h(0)=90$ [m], $v_{\rm P}(0)=20$ [m/s] and $v_{\rm F}(0)=28$ [m/s]. Both controllers can make the follower reach the uniform flow equilibrium within $20$ [s]. However, one can easily observe the performance difference in the transient phase. Panel (e) shows that the nonlinear controller initially generates a near-constant and comfortable desired acceleration $-a_{\rm com}$, which gradually decreases to $0$ when reaching the equilibrium.
Therefore, the desired speed $v_{\rm des}$ decreases linearly with respect to time and the follower speed tracks this desired speed closely as shown in panel (c). Due to moderate reactions, the tracking error $S$ remains less than $0.2$ [m/s] in the transient phase. Also, the distance $h$ quadratically decreases to the desired value $h_{\rm des}$ without overshoots/oscillations as shown in panel (a).
One may also observe that in this non-safety-critical scenario the magnitude of the collision-free feedforward term $a_{\rm cf}$ generates some disturbance that is less than $0.5$ [m/s$^{2}$]. However, this disturbance is compensated by the feedback term.

In contrast, the linear control strategy generates unexpected behaviors shown in Fig.~\ref{fig:sim_lead_slow_far}(b, d, f). Initially the distance error $\hat{h}$ dominates the control command on desired acceleration $a_{\rm des}$. Thus, the follower speeds up with extremely large acceleration to shorten the distance, and the desired speed $v_{\rm des}$ reaches $60$ [m/s] that is beyond tracking capability. Once the follower speed $v_{\rm F}$ is increased and the distance $h$ is shortened, the speed error $\hat{v}$ outweighs the distance error $\hat{h}$. Consequently, a harsh brake is applied that reaches $-6$ [m/s$^{2}$]. Due to overreaction, the tracking error $S $ is very large as shown in panel (d). Note that no saturation is applied to the desired acceleration or speed limit for the simulation here, but applying such saturations can not improve performance.

Fig.~\ref{fig:sim_lead_fast_far} depicts another scenario where the follower approaches a far-and-fast-moving predecessor with the initial conditions $h(0)=80$ [m], $v_{\rm P}(0)=20$ [m/s] and $v_{\rm F}(0)=16$ [m/s]. Similarly, one can observe the performance difference in the transient phase. Panel (e) shows that the nonlinear controller initially accelerates the follower to ensure that its speed $v_{\rm F}$ tracks the desired speed $v_{\rm des}$. This desired speed is slightly larger than the predecessor speed $v_{\rm P}$ such that the distance $h$ can be shortened gradually. Then the follower decelerates with a near-constant and comfortable acceleration $-a_{\rm com}$, which gradually decreases to $0$ at the final stage when reaching the equilibrium.
Panels (b, d, f) shows similar unexpected behaviors of the linear controller to those depicted in Fig.~\ref{fig:sim_lead_slow_far}.

Fig.~\ref{fig:sim_lead_slow_close} demonstrates a safety-critical scenario where the follower approaches a close-and-slow-moving predecessor with the initial conditions $h(0)=10$ [m], $v_{\rm P}(0)=20$ [m/s] and $v_{\rm F}(0)=25$ [m/s]. This is a representative scenario when a slow-moving vehicle cuts in and collision is imminent. Panels (a, c, e) illustrate the proper behaviors of the nonlinear controller. Specifically, the follower initially applies a harsh brake that reaches $-6$ [m/s$^{2}$]. The shortest inter-vehicle distance $h$ is around $7$ [m], which is larger than the designed safety distance $h_{\rm min}$. Note that the collision-free feedforward term $a_{\rm cf}$ generates around $-3$ [m/s$^{2}$] deceleration to avoid collision in this safety-critical scenario. Once the follower is slower than the predecessor and collision hazard disappears, the follower accelerates at a near-constant comfortable acceleration $a_{\rm com}$ to speed up while the distance still increases. Finally, the follower acceleration decreases to $0$ smoothly upon reaching the uniform flow equilibrium. Panels (b, d, f) shows unexpected behaviors of linear control strategy that the desired acceleration $a_{\rm des}$ exceeds the follower's braking capability at initial stage, and then reaches $1.5$ [m/s$^{2}$] due to over-braking.

Fig.~\ref{fig:sim_lead_fast_close} shows another scenario where the follower approaches a close-but-fast-moving predecessor with the initial conditions $h(0)=10$ [m], $v_{\rm P}(0)=20$ [m/s] and $v_{\rm F}(0)=16$ [m/s]. This is not a safety-critical scenario, but as shown in panels (b, d, f), the linear controller brakes at the initial stage because the distance error $\hat{h}$ outweighs the speed error $\hat{v}$, and then accelerates abruptly due to over-braking when the distance gets larger. However, the nonlinear controller provides reasonable performance indicated by panels (a, c, e). The follower speeds up with a near-constant acceleration $a_{\rm com}$ to catch up with the predecessor speed, while the distance still increases. When getting close to the uniform flow equilibrium, the follower acceleration decreases to $0$ and settles down at this desired equilibrium without overshoots/oscillations.

\begin{figure}[!t]
  \centering
  \includegraphics[scale=1]{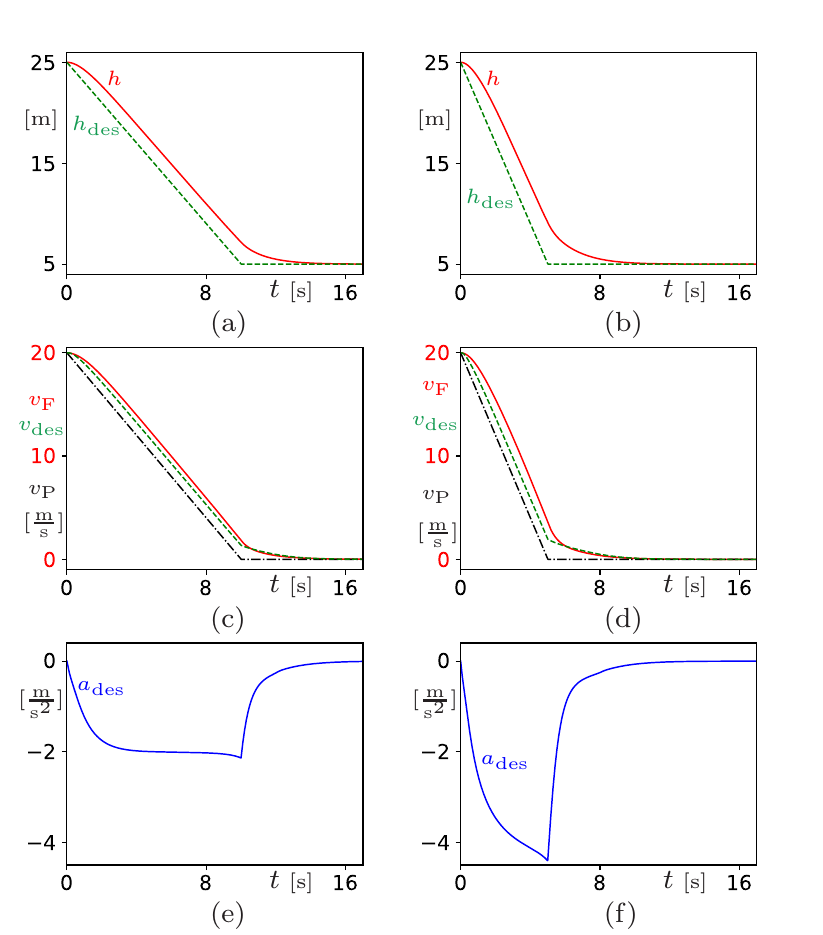}\\
  \caption{\label{fig:sim_lead_decel_stop}Performance of the nonlinear controller (\ref{eqn:ades_nonlin}-\ref{eqn:wrapper_q}) when the predecessor decelerates and stops completely. Left column: the predecessor decelerates at $-2$ [m/s$^{2}$]. Right column: the predecessor decelerates at $-4$ [m/s$^{2}$].}
\end{figure}

\subsection{Approaching Varying-speed Predecessor\label{sec:sim_vary_speed_lead}}
In this part we demonstrate the performance of the proposed nonlinear controller when predecessor speed varies. In Fig.~\ref{fig:sim_lead_decel_stop}, the left column and right column represent the scenarios where the predecessor decelerates to a complete stop with constant acceleration $-2$ [m/s$^{2}$] and $-4$ [m/s$^{2}$], respectively. The initial conditions are $h(0)=h_{\rm des}(0)=25$ [m] and $v_{\rm F}(0)=v_{\rm P}(0)=20$ [m/s]. The color scheme remains the same as that used in Figs.~\ref{fig:sim_lead_slow_far}-\ref{fig:sim_lead_fast_close}. One may observe the following: (1) the follower applies similar deceleration efforts as the predecessor, and completely stops at the standstill distance $h_{0}$; (2) the desired speed $v_{\rm des}$ decreases almost linearly with respect to time, and the follower speed keeps track of this desired speed; (3) the desired distance $h_{\rm des}$  decreases linearly with respect to time because the predecessor speed $v_{\rm P}$ decreases linearly with constant deceleration; (4) the actual distance $h$ also decreases linearly with respect to time before reaching the standstill distance $h_{0}$ because the speed error $\hat{v}$ remains almost constant; (5) the follower speed $v_{\rm F}$ and the actual distance $h$ are not exactly tracking the predecessor speed $v_{\rm P}$ or the desired distance $h_{\rm des}$, respectively. However, the follower speed $v_{\rm F}$ tracks the desired speed $v_{\rm des}$, implying that the resultant tracking error $S$ is rather small; cf.~(\ref{eqn:surf_S0}, \ref{eqn:surf_S}, \ref{eqn:surf_S0_interp}).

\begin{figure}[!t]
  \centering
  \includegraphics[scale=1]{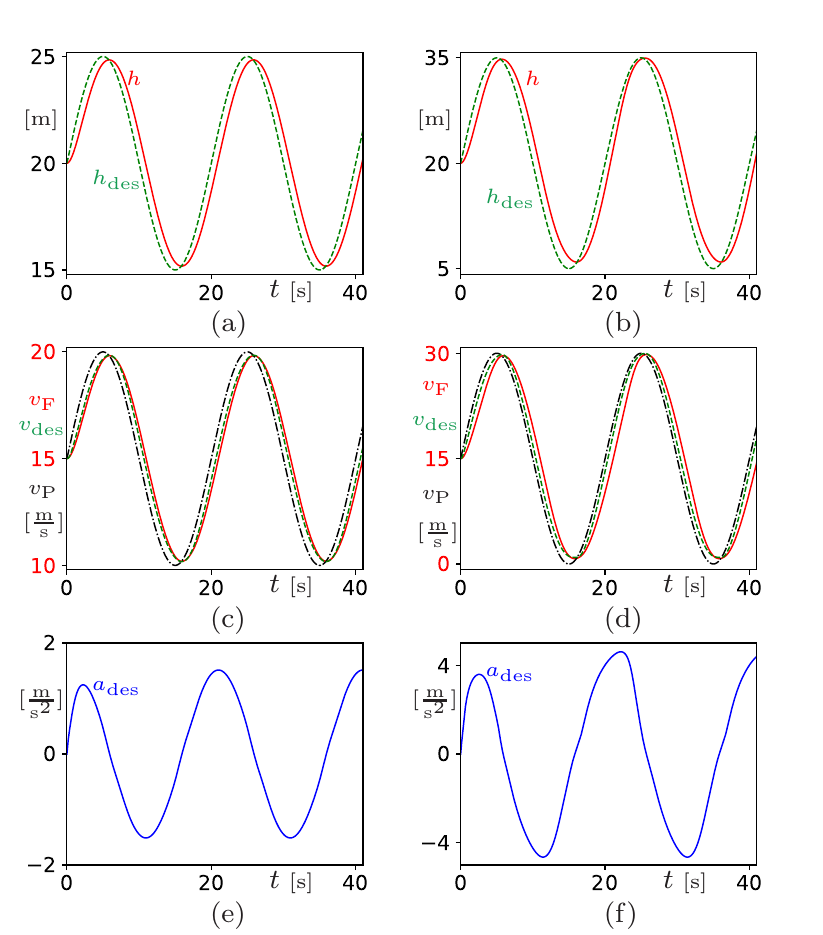}\\
  \caption{\label{fig:sim_lead_sin_osc}Performance of the nonlinear controller (\ref{eqn:ades_nonlin}-\ref{eqn:wrapper_q}) when the predecessor speed fluctuates around $v_{0}=15$ [m/s] at $f=0.05$ [Hz]. Left column: with amplitude $v_{\rm P}^{\rm amp} = 5$ [m/s]. Right column: with amplitude $v_{\rm P}^{\rm amp} = 15$ [m/s].}
\end{figure}

The simulation results above demonstrate that the proposed controller can settle down at the uniform flow equilibrium with proper acceleration effort in the transient phase when the follower state is far away from the equilibrium or this equilibrium changes significantly. This implies plant stability of the closed-loop system.
To maintain string stability, perturbations on predecessor speed must be attenuated by the follower. Indeed, one can apply realistic perturbations collected from field experiments to validate string stability, but it is difficult to conclude from simulation results. According to Fourier's theory, perturbations with finite energy can be approximated as sum of sinusoidal signals. Thus, we can impose a sinusoidal perturbation on the predecessor speed, that is,
\begin{align}
  v_{\rm P}(t) &= v_{0}+v_{\rm P}^{\rm amp}\sin(2\pi f t)\,.
\end{align}
Fig.~\ref{fig:sim_lead_sin_osc} shows one set of simulation results when the excitation frequency is $f=0.05$ [Hz], and the initial conditions are $h(0)=h_{\rm des}(0)=20$ [m] and $v_{\rm F}(0)=v_{\rm P}(0)=15$ [m/s]. The left panels show the results in a moderate scenario where $v_{\rm P}^{\rm amp} = 5$ [m/s], while the right panels illustrate the results in a more severe scenario where $v_{\rm P}^{\rm amp} = 15$ [m/s]. Observe that in both scenarios fluctuations on the follower speed are less than those imposed on the predecessor speed, implying that the system is string stable at $f=0.05$ [Hz]. We remark that the control gains used here satisfy the string stability conditions in Section~\ref{sec:analysis}. Thus, one will observe similar behaviors on fluctuation attenuation for other excitation frequencies, indicating that the system is string stable.

\subsection{Extensions to More Realistic Vehicle Models \label{sec:extension}}
So far we demonstrated the efficacy of the proposed nonlinear controller when the follower acceleration is capable of tracking the desired acceleration; cf.~\eqref{eqn:toy_model}. However, this assumption does not hold in practice. Based on physics, the longitudinal dynamics can be modeled as
\begin{equation}\label{eqn:long_dynamics}
  \begin{split}
    m_{\rm eff}\dot{v}_{\rm F} & =\dfrac{\eta\, T}{R} -m g \sin\phi -\mu m g \cos\phi-\rho(v_{\rm F}+v_{\rm w})^{2}\,,
  \end{split}
\end{equation}
by neglecting the flexibility of tires. Here, $m_{\rm eff}=m+\frac{J}{R^{2}}$ is the effective mass, containing the vehicle static mass $m$, the moment of inertia $J$ of rotating elements, and the wheel radius $R$. Also, $g$ is the gravitational constant, $\phi$ is the inclination angle, $\mu$ is the rolling resistance coefficient, $\rho$ is the air drag constant, $v_{\rm w}$ is the headwind speed, $\eta$ is the gear ratio, and $T$ is the actuation torque. In a more realistic scenario, the actuation torque is also governed by actuator dynamics, which is typically modeled as a first-order system, that is,
\begin{equation}\label{eqn:actuator_dyn}
  \dot{T} = -\dfrac{1}{\tau}\, T +\dfrac{1}{\tau}\, T_{\rm des}\,,
\end{equation}
where $T_{\rm des}$ is the desired torque, and $\tau$ is the time constant.

To ensure that the actual acceleration is able to track the desired acceleration, automated vehicles typically implement a low-level controller \cite{He_2019} in practice that calculates actuation torque based on desired acceleration and other vehicle parameters. In the following we demonstrate how the proposed controller can be integrated with these low-level controllers.
For simplicity, we also assume there is no headwind ($v_{\rm w}=0$) and the inclination angle $\phi$ can be obtained with onboard sensors (IMU+wheel-based acceleration measurements).

\subsubsection{Physics-based Model without Actuator Dynamics\label{sec:physics_model_no_actuator_dyn}}
First we consider the longitudinal dynamics \eqref{eqn:long_dynamics} and ignore the transmission dynamics \eqref{eqn:actuator_dyn} by assuming the actuation torque $T$ is capable of tracking desired torque $T_{\rm des}$ perfectly (i.e., $T=T_{\rm des}$).
One can apply feedback linearization technique to obtain the desired torque $T_{\rm des}$ based on acceleration command $u$ and nominal resistance. This yields the low-level controller
\begin{equation}\label{eqn:low_level_trq_ctrl}
T_{\rm des} =\dfrac{R}{\eta}(m_{\rm eff}\, u+m g \sin\phi+\bar{\mu} m g \cos\phi+\bar{\rho}\, v_{\rm F}^{2})\,,
\end{equation}
where $\bar{\mu}$ and $\bar{\rho}$ are the nominal values of rolling resistance coefficient and air drag constant, respectively. Integrating controller (\ref{eqn:low_level_trq_ctrl}) into dynamics \eqref{eqn:long_dynamics} leads to
\begin{equation}
    \dot{v}_{\rm F} = u +\Delta,
\end{equation}
where
\begin{equation}\label{eqn:disturbance}
\Delta =(\bar{\mu}-\mu) \dfrac{m}{m_{\rm eff}}g \cos\phi+\frac{\bar{\rho}-\rho}{m_{\rm eff}} v_{\rm F}^{2}
\end{equation}
is the disturbance of the model. Thus, the car-following dynamics can be simplified into
\begin{equation}\label{eqn:toy_model_disturb}
  \begin{split}
    \dot{h} & = v_{\rm P} -v_{\rm F}\,,\\
    \dot{v}_{\rm F} & = u+\Delta,
  \end{split}
\end{equation}
by choosing the distance $h$ and the speed $v_{\rm F}$ as state variables. Note that model \eqref{eqn:toy_model} is the case when the acceleration command $u$ is set to desired acceleration $a_{\rm des}$ directly in the absence of disturbance $\Delta$.
Typically this disturbance is small and can be handled with various control techniques. In the following, we utilize integral control technique as an example to extend the proposed car-following controller. That is,
\begin{equation}\label{eqn:ctrl_des_with_disturb}
  \begin{split}
    u & =a_{\rm des} +k_{\rm I}\, e \,,\\
    \dot{e} & = S\,,
  \end{split}
\end{equation}
where $a_{\rm des}$ is the controller given in (\ref{eqn:ades_nonlin}-\ref{eqn:wrapper_q}) and $S$ is the surface in \eqref{eqn:surf_S}. We remark that the feedback part in \eqref{eqn:ctrl_des_with_disturb} is a nonlinear proportional-integral controller on the surface $S$, which characterizes the error between range-dependent desired speed $v_{\rm des}$ and follower speed $v_{\rm F}$; cf.~(\ref{eqn:surf_S0_interp}-\ref{eqn:vdes_nonlin}). This controller ensures that the follower states evolve along the surface $S$ in the presence of disturbance while approaching and settling down around the desired equilibrium.
It is obvious that the closed-loop system (\ref{eqn:ades_nonlin}-\ref{eqn:wrapper_q}, \ref{eqn:toy_model_disturb}-\ref{eqn:ctrl_des_with_disturb}) possesses the desired uniform flow equilibrium. One can follow the procedure in Section~\ref{sec:analysis} to derive stability conditions.

\begin{figure}[!t]
  \centering
  \includegraphics[scale=1]{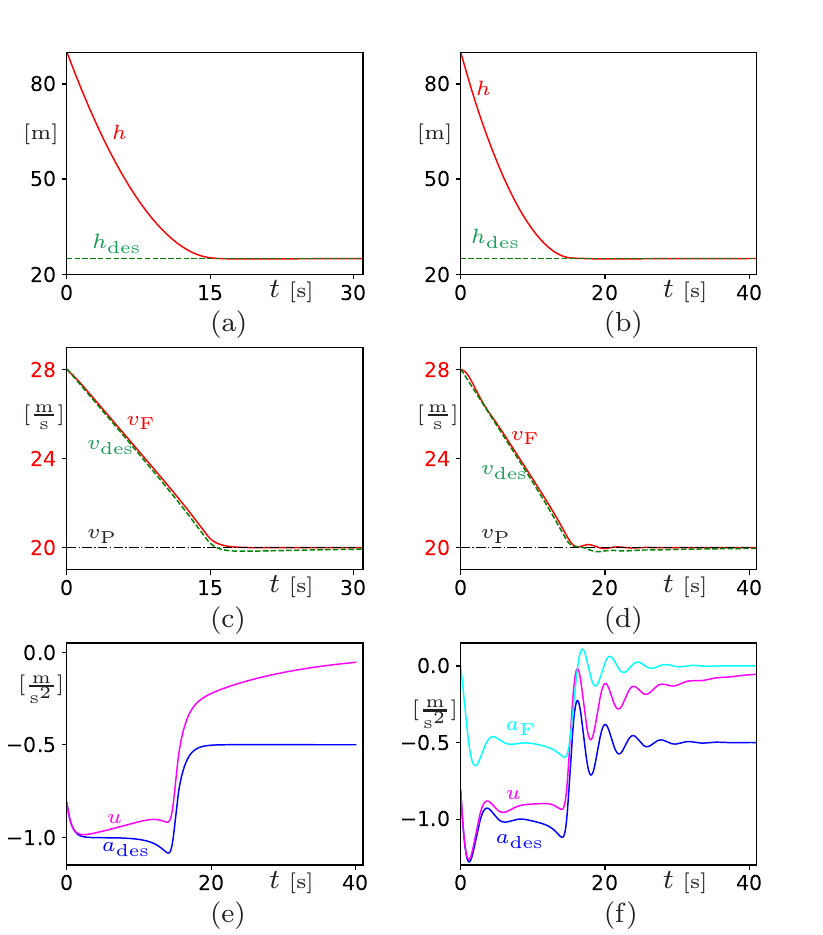}\\
  \caption{\label{fig:sim_lead_slow_far_disturb}The follower approaches a far-but-slow-moving predecessor. Left column: vehicle model \eqref{eqn:toy_model_disturb} with the nonlinear controller (\ref{eqn:ades_nonlin}-\ref{eqn:wrapper_q}, \ref{eqn:ctrl_des_with_disturb}). Right column: vehicle model \eqref{eqn:PT_model_disturb} with the nonlinear controller (\ref{eqn:ades_nonlin}-\ref{eqn:wrapper_q}, \ref{eqn:ctrl_des_with_disturb}).}
\end{figure}

In Fig.~\ref{fig:sim_lead_slow_far_disturb}, the left column demonstrates the response of the closed-loop system (\ref{eqn:ades_nonlin}-\ref{eqn:wrapper_q}, \ref{eqn:toy_model_disturb}-\ref{eqn:ctrl_des_with_disturb}) in the same scenario as that in Fig.~\ref{fig:sim_lead_slow_far}. The disturbance is $\Delta=0.5$ [m/s$^2$] and $k_{\rm I}=0.1$ [s$^{-1}$]; cf.~Table~\ref{tab:params}. One can observe that the follower applies a near-constant acceleration to decelerate while approaching the predecessor, and eventually settles down around the uniform-flow equilibrium. These transient behaviors are very similar to human-driving behaviors even in the presence of disturbance. Readers may simulate other scenarios as those in Fig.~\ref{fig:sim_lead_slow_far}-\ref{fig:sim_lead_sin_osc} with non-constant disturbances and notice that the car-following controller (\ref{eqn:ades_nonlin}-\ref{eqn:wrapper_q}) guarantees natural behaviors while approaching the equilibrium.

\subsubsection{Physics-based Model Including Actuator Dynamics\label{sec:physics_model_actuator_dyn}}
The same low-level controller \eqref{eqn:low_level_trq_ctrl} and extended car-following controller \eqref{eqn:ctrl_des_with_disturb} can be applied when the actuator dynamics \eqref{eqn:actuator_dyn} are considered along with the longitudinal dynamics \eqref{eqn:long_dynamics}.
We define the follower acceleration as
\begin{equation}\label{eqn:follower_accel}
  a_{\rm F} = \dot{v}_{\rm F}\,,
\end{equation}
and attempt to calculate its derivative such that the state variable can be transformed from actuator torque $T$ to follower acceleration $a_{\rm F}$.
By taking the derivative of \eqref{eqn:long_dynamics} and utilizing (\ref{eqn:long_dynamics}-\ref{eqn:low_level_trq_ctrl}, \ref{eqn:follower_accel}) in the substitution, we obtain
\begin{equation}\label{eqn:PT_model_disturb}
  \begin{split}
    \dot{h} & = v_{\rm P} -v_{\rm F}\,,\\
    \dot{v}_{\rm F} & = a_{\rm F}\,,\\
    \dot{a}_{\rm F} &= -\dfrac{1}{\tau}\, a_{\rm F} +\dfrac{1}{\tau}\, u +\dfrac{1}{\tau}\,\hat{\Delta}\,.
  \end{split}
\end{equation}
by choosing the distance $h$, the follower speed $v_{\rm F}$ and the follower acceleration $a_{\rm F}$ as state variables. The disturbance is
\begin{equation}
\hat{\Delta} = \Delta+\dfrac{m g\,\dot{\phi}\,\tau}{m_{\rm eff}}(\mu\sin\phi-\cos\phi)-2\rho\tau \dfrac{v_{\rm F}a_{\rm F}}{m_{\rm eff}}\,,
\end{equation}
where $\Delta$ is the same as that in \eqref{eqn:disturbance}. Again one can verify that the closed-loop system (\ref{eqn:ades_nonlin}-\ref{eqn:wrapper_q}, \ref{eqn:ctrl_des_with_disturb}, \ref{eqn:PT_model_disturb}) possesses the desired uniform flow equilibrium, and can also derive stability conditions following the procedures in Section~\ref{sec:analysis}.

The right column of Fig.~\ref{fig:sim_lead_slow_far_disturb} illustrates the response of the system \eqref{eqn:PT_model_disturb} with the controller (\ref{eqn:ades_nonlin}-\ref{eqn:wrapper_q}, \ref{eqn:ctrl_des_with_disturb}) using the same scenario as the left column. The time constant $\tau=0.8$ [s], the disturbance $\hat{\Delta}=0.5$ [m/s$^2$] and $k_{\rm I}=0.1$ [s$^{-1}$]. One can observe similar natural driving behaviors when the follower decelerates at near-constant acceleration to approach the predecessor. However, due to the inertial drag in actuator dynamics, the simple integral controller in \eqref{eqn:ctrl_des_with_disturb} generates some oscillations with amplitude around $0.1$ [m/s$^{2}$] while it attempts to compensate the disturbance. We remark that as the inertial drag increases, these oscillations will gradually become noticeable, and eventually there will be no gains that can stabilize the system using the extended controller \eqref{eqn:ctrl_des_with_disturb}. In such cases one may need to apply other techniques to extend the car-following controller \eqref{eqn:ades_nonlin} that guarantees the natural driving behaviors. This will be discussed further in a follow-up paper.

\subsection{Summary\label{sec:summary}}
In this section, we compared simulation results of the proposed nonlinear controller against the widely-used linear controller, and demonstrated its effectiveness and advantages.
Firstly, it preserves stability results of linear controller in the neighborhood of the uniform flow equilibrium. In other words, the steady state response remains unchanged due to topological equivalence around that equilibrium.
Secondly, the proposed controller applies a near-constant acceleration strategy that mimics human-driving behavior while approaching the equilibrium with large initial errors. This natural behavior improves passenger comfort significantly in the transient phase.
Moreover, the improvement in transient response can enhance safety in case of temporary failures/malfunctions as discussed in the introduction. Last but not least, the proposed controller preserves simplicity that only requires constant time complexity and constant space complexity in implementation.

We also demonstrated the extended format of the proposed car-following controller for more realistic models when disturbances and actuator dynamics are considered. The proposed car-following controller (\ref{eqn:ades_nonlin}-\ref{eqn:wrapper_q}) serves as the baseline design that ensures natural driving behaviors to increase passenger comfort, while the extended part guarantees reasonable acceleration tracking in the presence of disturbance and dynamics. This design separates physics-based but model-free design of natural driving behaviors, and the model-dependent design of acceleration tracking.

\section{Conclusion\label{sec:conclusion}}

In this paper, we proposed a nonlinear car-following controller that considers the transient response in the approaching phase, the steady-state response around the uniform flow equilibrium, and safety-critical scenario for collision-avoidance. We studied plant stability, string stability and tracking performance of the controller, and provided conditions and guidelines on the selection of control gains. We also used simulations to demonstrate that the proposed controller provides satisfactory performance in different scenarios. In contrast, we showed that the widely-used linear car-following controller generates unexpected behaviors especially in the transient phase before reaching the uniform flow equilibrium. Future research directions may include the integration of actuator dynamics, consideration of inherent time delays, incorporation of measurement imperfections, extension to connected automated vehicle systems, analysis on nonlinear dynamics, etc.




\begin{IEEEbiography}
[{\includegraphics[width=1in,height=1.25in,clip,keepaspectratio]{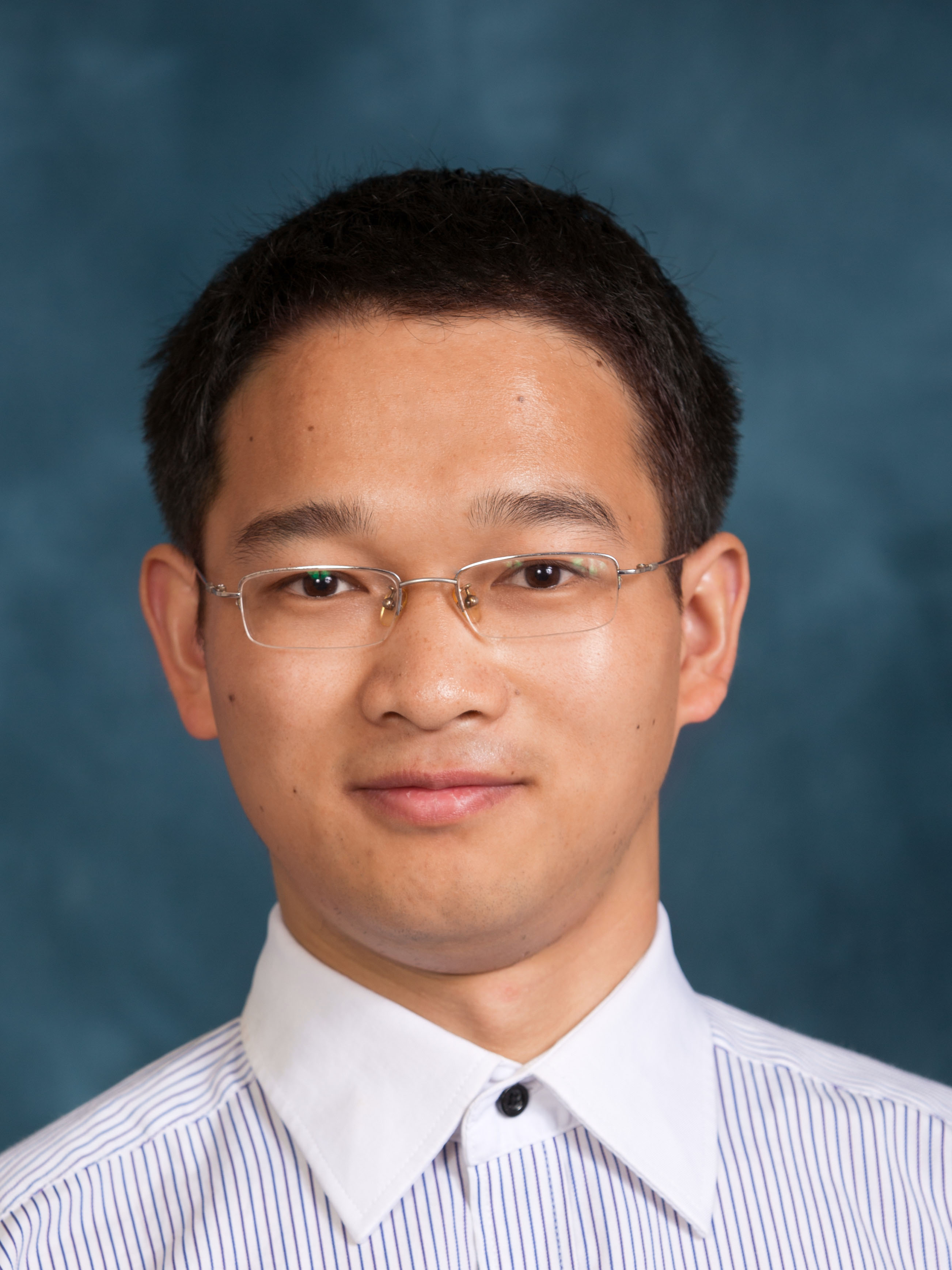}}]{Wubing B. Qin}
received his BEng degree in School of Mechanical Science and Engineering from
Huazhong University of Science and Technology, China in 2011, and his MSc degree and PhD degree in
Mechanical Engineering from the University of Michigan, Ann Arbor in 2016 and 2018, respectively.
Curretly he is a research engineer at Ford Motor Company. His
research focuses on dynamics and control of connected automated vehicles, digital systems, ground robotics,
and nonlinear and stochastic systems with time delays.
\end{IEEEbiography}

\end{document}